\documentclass[aps,showpacs,preprintnumbers,nofootinbib]{revtex4}
\usepackage{amssymb}
\usepackage{amsmath}
\usepackage{bm}

\input{tcilatex}

\begin{document}

\title{Modified Brans-Dicke theory of gravity from five-dimensional vacuum}
\author{$^{a}$Jos\'{e} Edgar Madriz Aguilar\thanks{%
E-mail address: jemadriz@fisica.ufpb.br}, $^{a}$Carlos Romero \thanks{%
E-mail address: cromero@fisica.ufpb.br} and $^{b}$Adriano Barros}
\affiliation{$^{a}$Departamento de F\'{\i}sica, Universidade Federal da Para\'{\i}ba,
Caixa Postal 5008, 58059-970 \\
Jo\~{a}o Pessoa, PB, Brazil.\\
$^{b}$Departamento de F\'{\i}sica, Universidade Federal de Roraima,
69310-270, Boa Vista, RR, Brazil\\
E-mail: jemadriz@fisica.ufpb.br; cromero@fisica.ufpb.br}

\begin{abstract}
We investigate, in the context of five-dimensional (5D) Brans-Dicke theory
of gravity, the idea that macroscopic matter configurations can be generated
from pure vacuum in five dimensions, an approach first proposed in the
framework of general relativity. We show that the 5D Brans-Dicke vacuum
equations when reduced to four dimensions lead to a modified version of
Brans-Dicke theory in four dimensions (4D). As an application of the
formalism, we obtain two five-dimensional extensions of four-dimensional 
O'Hanlon and Tupper vacuum solution and show that they lead two different
cosmological scenarios in 4D.
\end{abstract}

\maketitle

\vskip .5cm


\section{Introduction}

Recourse to extra dimensions, beyond space and time, has been considered
during the last decades as probably the best if not the only way to achieve
unification of gravity with the other interactions of physics. This idea,
which goes back to the works of G. Nordstr\"{o}m \cite{Nordstrom} and to the
original form of Kaluza-Klein theory \cite{Kaluza-Klein}, has been the
source of an enormous amount of research that have ultimately culminated in
string theory and its recent generalization, the so-called brane theory \cite%
{Pavsic}. Parallelly to this development there has recently been a fair deal
of work on a proposal by P. Wesson and collaborators, known as the
induced-matter theory (IMT), a non-compactified approach to
five-dimensional\ Kaluza-Klein gravity \cite{Wesson}. The basic principle of
the IMT approach is that classical macroscopic properties of physical
quantities, such as matter density and pressure, may be given a geometrical
interpretation if one assumes the existence of a fundamental
five-dimensional\ space $M^{5}$, in which our usual four-dimensional (4D)
spacetime is embedded. It is also assumed that the metric $g_{AB}$ of $M^{5}$
is a solution of the vacuum Einstein field equations in five dimensions (5D) 
\footnote{%
Throughout capital Latin indices take value in the range (0,1,...4) while
Greek indices run from (0,1,2,3). We shall also denote the fifth coordinate $%
y^{4}$ by $l$ and the first four coordinates $y^{\mu }$\ (the "spacetime" 
\textit{\ }coordinates) by $x^{\mu }$, that is, $y^{A}=(x^{\mu },l)$, with $%
\mu =0,1,...,3.)$}%
\begin{equation*}
R_{AB}=0.
\end{equation*}%
The main idea of IMT theory is that the above equation may be re-expressed
so that in 4D, the four-dimensional components of $g_{AB}$, which we denote
by $g_{\alpha \beta }$, satisfy Einstein's equations 
\begin{equation*}
G_{\alpha \beta }=8\pi GT_{\alpha \beta }
\end{equation*}%
in which\ the "energy-momentum tensor" $T_{\alpha \beta }$ has a purely
geometrical origin and this fact is interpreted as saying that
four-dimensional matter is induced from geometry in five-dimensional space 
\cite{Wesson}. At the time the first papers on IMT were published there was
no guarantee that any energy-momentum tensor could be induced in this way.
Mathematically rephrased the question then was: Is it possible to
isometrically embed any solution of Einstein\'{}s equations in a 5D
Ricci-flat space? The anwer to this question was provided by the
Campbell-Magaard theorem, which asserts that any analytic $n$-dimensional
Riemannian space can be locally embedded in a $(n+1)$-dimensional Ricci-flat
space \cite{Campbell,Romero,Seahra}. The rediscovery of Campbell-Magaard
theorem by physicists and its connection with modern spacetime embedding
theories gave rise to a series of new results and extensions in the cases
when the ambient space: a) is an Einstein space \cite{Dahia1}, b) has a
nondegenerate Ricci-tensor \cite{Dahia2}, c) is sourced by scalar-fields 
\cite{Dahia3}. The latter case includes embeddings in spaces that are
solutions of the vacuum Brans-Dicke field equations.

The Brans-Dicke theory of gravity represents a natural extension of General
Relativity where Newton's gravitational constant is replaced by a
non-minimally coupled scalar field \cite{Brans-Dicke}. Brans-Dicke theory
or, more generally, scalar-tensor theories in higher dimensions have also
been studied in different contexts. For instance, cosmological questions
related to quintessence, inflation, acceleration of the Universe, and the
so-called braneworld scenario have been examined in the framework of
higher-dimensional scalar-tensor theories \cite{BDfiveD}. Certainly, an
attractive feature of the Brans-Dicke theory concerns the geometrical nature
of the scalar field, as opposed to other models in which the scalar field
has to be separately postulated and thereby have an \textit{ad hoc }status.

In this paper we investigate a modified version of the induced-matter theory
in which the role of General Relativity as the fundamental underlying theory
is replaced by the Brans-Dicke theory of gravity. As we shall see, the
five-dimensional Brans-Dicke vacuum equations when reduced to four
dimensions lead to a modified 4D\ Brans-Dicke theory in which matter can
still be viewed as generated by the fifth dimension.

The paper is organized as follows. In Section II we set up the Brans-Dicke
theory in five dimensions and write the field equations in the absence of
matter, a configuration to which we refer as the 5D vacuum state. We then
proceed to Section III, where we show that the 5D vacuum field equations
lead to a modified version of four-dimensional Brans-Dicke theory. In
Sections IV and V we illustrate the formalism developed in the previous
sections by extending the O'Hanlon and Tupper vacuum solution to five
dimensions in two different ways and show that when viewed form a 4D
perspective they lead to different cosmological scenarios. We conclude with
some remarks in Section VI.

\section{Brans-Dicke field equations in five dimensions}

In order to describe the Brans-Dicke (BD) theory of gravity in 5D in a
general manner, we consider the 5D action 
\begin{equation}
\mathcal{S}[g_{AB},\varphi ]=\frac{1}{16\pi }\int \sqrt{\left\vert \frac{%
^{(5)}g}{^{(5)}g_{0}}\right\vert }\left[ \varphi ^{(5)}\mathcal{R}+L_{matt}-%
\frac{\omega }{\varphi }g^{AB}\varphi _{,A}\varphi _{,B}\right] d^{5}x,
\label{f1}
\end{equation}%
where $\varphi $ is a scalar field that describes the gravitational coupling
in 5D, $^{(5)}\mathcal{R}$ is the 5D Ricci scalar, $^{(5)}g$ is the
determinant of the 5D metric here denoted by $g_{AB}$, $^{(5)}g_{0}$ is a
dimensionalization constant, $L_{matt}$ represents the matter lagrangian and 
$\omega $ is a constant parameter.\newline
\qquad The field equations obtained from the action (\ref{f1}) are 
\begin{eqnarray}
&&G_{AB}=\frac{8\pi }{\varphi }T_{AB}+\frac{\omega ^{2}}{\varphi ^{2}}\left[
\varphi _{,A}\varphi _{,B}-\frac{1}{2}g_{AB}\varphi ^{,C}\varphi _{,C}\right]
+\frac{1}{\varphi }\left[ \varphi _{;A;B}-g_{AB}\varphi ^{;C}\,_{;C}\right] ,
\label{f2} \\
&&\frac{2\omega }{\varphi }\varphi ^{;C}\,_{;C}-\frac{\omega }{\varphi ^{2}}%
g^{AB}\varphi _{,A}\varphi _{,B}+^{(5)}\mathcal{R}=0,  \label{f3}
\end{eqnarray}%
with $(;)$ denoting covariant derivative, $T_{AB}$ being the energy-momentum
tensor of matter and $G_{AB}=R_{AB}-(1/2)\,^{(5)}\mathcal{R}g_{AB}$ is the
Einstein tensor in five dimensions. Taking the trace of the expression ( \ref%
{f2}) gives 
\begin{equation}
^{(5)}\mathcal{R}=-\frac{16\pi }{3\varphi }T+\frac{\omega }{\varphi ^{2}}%
\varphi ^{;C}\varphi _{;C}+\frac{8}{3\varphi }\varphi ^{;C}\,_{;C}\,,
\label{f4}
\end{equation}%
where $T=T_{A}^{A}$. By substituting (\ref{f4}) in (\ref{f3}) one has 
\begin{equation}
^{(5)}\Box \varphi =\frac{8\pi }{4+3\omega }T,  \label{f5}
\end{equation}%
where $^{(5)}\Box \varphi =\varphi ^{;A}\,_{;A}$. Since (\ref{f3}) and (\ref%
{f5}) are not independent, then we may consider (\ref{f2}) and (\ref{f5}) as
the 5D field equations.

Let us now consider the 5D vacuum case. We define the 5D vacuum state in BD
theory as a configuration in which $T_{AB}=0$, i.e. the 5D vacuum is defined
as the absence of matter in 5D. This is equivalent to setting $L_{matt}=0$
in the action (\ref{f1}). Thus, the 5D vacuum field equations will be given
by 
\begin{eqnarray}
G_{AB} &=&\frac{\omega ^{2}}{\varphi ^{2}}\left[ \varphi _{,A}\varphi _{,B}-%
\frac{1}{2}g_{AB}\varphi ^{,C}\varphi _{,C}\right] +\frac{1}{\varphi }\left[
\varphi _{;A;B}-g_{AB}\varphi ^{;C}\,_{;C}\right] ,  \label{f6} \\
^{(5)}\Box \varphi &=&0.  \label{f7}
\end{eqnarray}%
while the equation (\ref{f4}) in vacuum becomes 
\begin{equation}
^{(5)}R=\frac{\omega }{\varphi ^{2}}g^{AB}\varphi _{,A}\varphi _{,B},
\label{f8}
\end{equation}%
which implies that in vacuum the scalar curvature is generated only by the
free scalar field $\varphi $.

\section{The effective four-dimensional modified Brans-Dicke theory}

Following the idea, as in the IMT approach, that matter can be generated
from five-dimensional pure vacuum, we investigate the possibility of
obtaining the Brans-Dicke four-dimensional field equations with matter as a
subset of the field equations of a Brans-Dicke field theory in 5D vacuum.
Our approach, in fact, will be more general in the sense that we shall deal
with a somewhat modified version Brans-Dicke theory in 4D, that is, one in
which a scalar potential $V(\varphi )$ is taken into account. In a
cosmological setting it is well known that the scalar field in 4D
Brans-Dicke theory can lead to cosmological acceleration only when the
coupling parameter $\omega $ varies with time \cite{Banerjee} or when there
is a scalar potential which usually is added by means of an \textit{ad hoc }%
assumption \cite{Sen}. A scalar field obtained in this way has the desired
property that it accelerates the universe, thereby leading to quintessence
or K-essence. At this point we would like to call attention for the fact
that in the formalism developed in the present paper a potential exhibiting
the mentioned property can be geometrically induced due to the presence of
the fifth coordinate, avoiding \textit{ipso facto} its introduction as "by
hand". In fact, the idea of inducing 4D scalar potentials $V(\varphi )$
geometrically from a 5D space-time has already been introduced in \cite{Anab}%
, in the context of inflationary models. Let us now develop a formalism
which, in a way, generalises the induced-matter approach and which leads to
\ a slightly modified Brans-Dicke theory in four dimensions, these
modifications being caused essentially by the presence of the scalar
potential.

We start by considering the 5D line element written in local coordinates $%
y^{A}=(x^{\mu },l)$ 
\begin{equation}
dS^{2}=g_{AB}dx^{A}dx^{B}=g_{\mu \nu }(x^{A})dx^{\mu }dx^{\nu
}+g_{44}(x^{A})(dl)^{2}.  \label{ef1}
\end{equation}%
Then we assume that the 5D space is foliated by a family of hypersurfaces $%
\Sigma $ defined by $l=const$. The metric induced on a generic hypersurface $%
\Sigma _{0}$ $(l=l_{0})$ has the form 
\begin{equation}
ds^{2}=g_{\mu \nu }(x^{\alpha },l_{0})dx^{\mu }dx^{\nu }.  \label{4dmetric}
\end{equation}%
On the other hand, the action (\ref{f1}) on $\Sigma _{0}$ takes the form 
\begin{equation}
S_{\Sigma _{0}}=\frac{1}{16\pi }\int \sqrt{|\,^{(4)}g|}\left[ \varphi
\,^{(4)}R+L_{IM}-\frac{\omega }{\varphi }g^{\alpha \beta }\varphi _{,\alpha
}\varphi _{,\beta }+V(\varphi )\right] d^{4}x,  \label{ag1}
\end{equation}%
where $L_{IM}$ is the Lagrangian corresponding to the induced matter. Note
that $L_{IM}$ comes from the first term in action (\ref{f1}) when it is
splitted in a 4D part plus a fifth coordinate contribution and finally
evaluated on $\Sigma _{0}$. Here we have made the formal identification 
\begin{equation}
V(\varphi )=-\left. \frac{\omega }{\varphi }g^{44}\varphi _{,4}\varphi
_{,4}\right\vert _{\Sigma _{0}}.  \label{potential}
\end{equation}%
At a first view, $V(\varphi )$ looks like a kinetic term instead of a
potential. However, as it has been shown in \cite{Anab}, depending on the
metric background and considering separable scalar fields such an
identification is valid. \newline

By taking the space-time component ($A=\alpha $, $B=\beta $) of the 5D
Brans-Dicke field equations (\ref{f6}) we obtain 
\begin{eqnarray}
\hat{G}_{\alpha \beta } &=&\frac{\omega }{\varphi ^{2}}\left[ \varphi
_{,\alpha }\varphi _{,\beta }-\frac{1}{2}g_{\alpha \beta }\varphi ^{,\sigma
}\varphi _{,\sigma }\right] +\frac{1}{\varphi }\left[ \varphi _{;\alpha
;\beta }-g_{\alpha \beta }(^{(4)}\Box \varphi -\frac{1}{2}V(\varphi ))\right]
-  \notag  \label{ef2} \\
&&-\frac{g_{\alpha \beta }}{\varphi }\left[ g^{44}\overset{\star \star }{%
\varphi }+\frac{1}{2}\left( g^{\sigma \rho }\overset{\star }{g}_{\rho \sigma
}g^{44}+(g^{44})^{2}\overset{\star }{g}_{44}\right) \overset{\star }{\varphi 
}+\frac{1}{2}g^{44}g_{44,\sigma }\varphi ^{,\sigma }\right]
\end{eqnarray}%
where here $(^{\wedge })$ denotes the 4D part of a 5D quantity, the
semicolon indicates covariant derivative, the star stands for derivative
with respect to the fifth coordinate $l$ and $^{(4)}\Box \varphi =\varphi
^{;\sigma }\,_{;\sigma }$ denotes the four-dimensional D'Alembertian
operator acting on the field $\varphi $. On the other hand, it can directly
be shown that \cite{Ponce} 
\begin{equation}
\hat{G}_{\alpha \beta }=G_{\alpha \beta }-\frac{8\pi }{\varphi }T_{\alpha
\beta }^{(IMT)},  \label{ef3}
\end{equation}%
where $T_{\alpha \beta }^{(IMT)}$ \ is given by 
\begin{equation}
\frac{8\pi }{\varphi }T_{\alpha \beta }^{(IMT)}=g^{44}(g_{44,\alpha
})_{;\beta }-\frac{1}{2}(g^{44})^{2}\left[ g^{44}\overset{\star }{g}_{44}%
\overset{\star }{g}_{\alpha \beta }-\overset{\star \star }{g}_{\alpha \beta
}+g^{\lambda \mu }\overset{\star }{g}_{\alpha \lambda }\overset{\star }{g}%
_{\beta \mu }-\frac{1}{2}g^{\mu \nu }\overset{\star }{g}_{\mu \nu }\overset{%
\star }{g}_{\alpha \beta }+\frac{1}{4}g_{\alpha \beta }\left( \overset{\star 
}{g}^{\mu \nu }\overset{\star }{g}_{\mu \nu }+(g^{\mu \nu }\overset{\star }{g%
}_{\mu \nu })^{2}\right) \right] .  \label{ef4}
\end{equation}%
Inserting (\ref{ef3}) into (\ref{ef2}) we obtain 
\begin{equation}
G_{\alpha \beta }=\frac{8\pi }{\varphi }T_{\alpha \beta }^{(BD)}+\frac{%
\omega }{\varphi ^{2}}\left[ \varphi _{,\alpha }\varphi _{,\beta }-\frac{1}{2%
}g_{\alpha \beta }\varphi ^{,\sigma }\varphi _{,\sigma }\right] +\frac{1}{%
\varphi }\left[ \varphi _{;\alpha ;\beta }-g_{\alpha \beta }(^{(4)}\Box
\varphi -\frac{1}{2}V(\varphi ))\right] ,  \label{ef5}
\end{equation}%
where $T_{\alpha \beta }^{(BD)}$ is the induced energy-momentum tensor on
the effective 4D modified Brans-Dicke theory, which have the form 
\begin{equation}
8\pi T_{\alpha \beta }^{(BD)}=8\pi T_{\alpha \beta }^{(IMT)}-g_{\alpha \beta
}\left[ g^{44}\overset{\star \star }{\varphi }+\left( \frac{1}{2}g^{\sigma
\rho }\overset{\star }{g}_{\rho \sigma }+\frac{1}{2}(g^{44})^{2}\overset{%
\star }{g}_{44}\right) \overset{\star }{\varphi }+\frac{1}{2}%
g^{44}g_{44,\sigma }\varphi ^{,\sigma }\right] .  \label{ef6}
\end{equation}%
It is important to note that in comparison with the IMT in our case the
induced energy-momentum tensor is composed of two parts. The first part is
induced from the extra-dimensional part of the metric and it is the same as
in the IMT approach. The second part adds a new contribution that depends on
the scalar field $\varphi $\ and its derivatives with respect to the fifth
coordinate.

Now setting $A=4$ and $B=\mu $ in equation (\ref{f6}) yields 
\begin{equation}
G_{4\mu }=\frac{\omega ^{2}}{\varphi ^{2}}\overset{\star }{\varphi }\varphi
_{,\mu }+\frac{\overset{\star }{\varphi }_{,\mu }}{\varphi }-\frac{1}{%
2\varphi }\left[ g^{\alpha \beta }\overset{\star }{g}_{\beta \mu }\varphi
_{,\alpha }-g^{44}g_{44,\mu }\overset{\star }{\varphi }\right] .  \label{ef7}
\end{equation}%
From the IMT we know that $G_{4\mu }$ can be expressed as $G_{4\mu }=\sqrt{%
g_{44}}P^{\beta }\,_{\mu ;\beta }$, where $P_{\alpha \beta }=[1/(2\sqrt{%
g_{44}})](\overset{\star }{g}_{\alpha \beta }-g_{\alpha \beta }g^{\mu \nu }%
\overset{\star }{g}_{\mu \nu })$ is a conserved quantity. Therefore,
equation (\ref{ef7}) becomes 
\begin{equation}
\frac{\omega ^{2}}{\varphi ^{2}}\overset{\star }{\varphi }\varphi _{,\mu }+%
\frac{\overset{\star }{\varphi }_{,\mu }}{\varphi }-\frac{1}{2\varphi }\left[
g^{\alpha \beta }\overset{\star }{g}_{\beta \mu }\varphi _{,\alpha
}-g^{44}g_{44,\mu }\overset{\star }{\varphi }\right] -\sqrt{g_{44}}P^{\beta
}\,_{\mu ;\beta }=0.  \label{ef8}
\end{equation}%
The last component $A=4$, $B=4$ of (\ref{f6}) reads 
\begin{equation}
G_{44}=\frac{\omega ^{2}}{\varphi ^{2}}\left[ \overset{\star }{\varphi }^{2}-%
\frac{1}{2}g_{44}\left( \varphi ^{,\alpha }\varphi _{,\alpha }+g^{44}\overset%
{\star }{\varphi }^{2}\right) \right] -\frac{1}{\varphi }\varphi _{;\alpha
}^{;\alpha }-\frac{1}{2\varphi }g^{\sigma \rho }\overset{\star }{g}_{\sigma
\rho }\overset{\star }{\varphi },  \label{ef9}
\end{equation}%
where $G_{44}=R_{44}-(1/2)g_{44}\,^{(5)}R$. On the other hand $R_{44}$ is
given by 
\begin{equation}
R_{44}=-g_{44}g^{\mu \nu }(g_{44,\mu })_{;\nu }-\frac{1}{2}\overset{\star }{g%
}^{\lambda \beta }\overset{\star }{g}_{\lambda \beta }-\frac{1}{2}g^{\lambda
\beta }\overset{\star \star }{g}_{\lambda \beta }+\frac{1}{2}g^{44}\overset{%
\star }{g}_{44}g^{\lambda \beta }\overset{\star }{g}_{\lambda \beta }-\frac{1%
}{4}g^{\mu \beta }g^{\lambda \sigma }\overset{\star }{g}_{\lambda \beta }%
\overset{\star }{g}_{\mu \sigma },  \label{ef10}
\end{equation}%
while the 5D scalar curvature is 
\begin{equation}
^{(5)}R=\,^{(4)}R-\frac{1}{4}(g^{44})^{2}\left[ \overset{\star }{g}^{\mu \nu
}\overset{\star }{g}_{\mu \nu }+\left( g^{\mu \nu }\overset{\star }{g}_{\mu
\nu }\right) ^{2}\right] .  \label{ef11}
\end{equation}%
Thus, the 5D component $G_{44}$ becomes 
\begin{eqnarray}
G_{44} &=&-g_{44}g^{\mu \nu }\left( g_{44,\mu }\right) _{;\nu }-\frac{1}{2}%
\overset{\star }{g}^{\lambda \beta }\overset{\star }{g}_{\lambda \beta }-%
\frac{1}{2}g^{\lambda \beta }\overset{\star \star }{g}_{\lambda \beta }+%
\frac{1}{2}g^{44}\overset{\star }{g}_{44}g^{\lambda \beta }\overset{\star }{g%
}_{\lambda \beta }-\frac{1}{4}g^{\mu \beta }g^{\lambda \sigma }\overset{%
\star }{g}_{\lambda \beta }\overset{\star }{g}_{\mu \sigma }-  \notag \\
&-&\frac{1}{2}g_{44}\left[ ^{(4)}R-\frac{1}{4}(g^{44})^{2}\left( \overset{%
\star }{g}^{\mu \nu }\overset{\star }{g}_{\mu \nu }+(g^{\mu \nu }\overset{%
\star }{g}_{\mu \nu })^{2}\right) \right]  \label{ef12}
\end{eqnarray}%
From (\ref{ef12}) and (\ref{ef9}) we have 
\begin{eqnarray}
&&-g_{44}g^{\mu \nu }(g_{44,\mu })_{;\nu }-\frac{1}{2}\overset{\star }{g}%
^{\lambda \beta }\overset{\star }{g}_{\lambda \beta }-\frac{1}{2}g^{\lambda
\beta }\overset{\star \star }{g}_{\lambda \beta }+\frac{1}{2}g^{44}\overset{%
\star }{g}_{44}g^{\lambda \beta }\overset{\star }{g}_{\lambda \beta }-\frac{1%
}{4}g^{\mu \beta }g^{\lambda \sigma }\overset{\star }{g}_{\lambda \beta }%
\overset{\star }{g}_{\mu \sigma }-\frac{1}{2}g_{44}^{(4)}R+  \notag
\label{ef13} \\
&&+\frac{1}{8}g^{44}[\overset{\star }{g}^{\mu \nu }\overset{\star }{g}_{\mu
\nu }+(g^{\mu \nu }\overset{\star }{g}_{\mu \nu })^{2}]=\frac{\omega ^{2}}{%
\varphi ^{2}}\left[ \overset{\star }{\varphi }^{2}-\frac{1}{2}g^{44}(\varphi
^{,\alpha }\varphi _{,\alpha }+g^{44}\overset{\star }{\varphi }^{2})\right] -%
\frac{1}{\varphi }\varphi ^{;\alpha }\,_{;\alpha }-\frac{1}{2\varphi }%
g^{\sigma \rho }\overset{\star }{g}_{\sigma \rho }\overset{\star }{\varphi }.
\end{eqnarray}%
We thus see that equation (\ref{f6}) is equivalent to the system of
equations (\ref{ef5}), (\ref{ef9}) and (\ref{ef13}). On the other hand,
equation (\ref{f7}) evaluated on the hypersurface $\Sigma _{0}$ can be
written as 
\begin{equation}
\frac{2\omega }{\varphi }\,^{(4)}\Box \varphi -\frac{\omega }{\varphi ^{2}}%
g^{\alpha \beta }\varphi _{,\alpha }\varphi _{,\beta }+V^{\prime }(\varphi
)+\,^{(4)}\!R=0  \label{ef14}
\end{equation}%
where we have made the formal identification 
\begin{equation}
V^{\prime }(\varphi )\equiv \left. \left[ \frac{2\omega }{\varphi }\frac{1}{%
\sqrt{|\,^{(5)}g|}}\frac{\partial }{\partial l}\left( \sqrt{|\,^{(5)}g|}%
g^{44}\varphi _{,4}\right) -\frac{\omega }{\varphi ^{2}}g^{44}\varphi
_{,4}\varphi _{,4}\right] \right\vert _{\Sigma _{0}}  \label{ef15}
\end{equation}%
It is clear that equations (\ref{ef5}) and (\ref{ef14}) are a subset of the
field equations in 5D vacuum (\ref{f6}) and (\ref{f3}), while by taking the
trace of (\ref{ef5}) we have 
\begin{equation}
^{(4)}\!R=-\frac{8\pi }{\varphi }T^{(BD)}+\frac{\omega }{\varphi ^{2}}%
\varphi ^{,\alpha }\varphi _{,\alpha }+\frac{1}{\varphi }\left[
3\,^{(4)}\Box \varphi -2V(\varphi )\right] ,  \label{ef16}
\end{equation}%
with $T^{(BD)}=g^{\alpha \beta }T_{\alpha \beta }^{(BD)}$. Now if we
susbtitute (\ref{ef16}) into (\ref{ef14}) we get 
\begin{equation}
^{(4)}\!\Box \varphi =\frac{8\pi }{3+2\omega }T^{(BD)}+\frac{2}{3+2\omega }%
V(\varphi )-\frac{\varphi }{3+2\omega }V^{\prime }(\varphi ),  \label{ef17}
\end{equation}%
It is now clear that (\ref{ef5}) and (\ref{ef17}) correspond to the field
equations of a more general version of four-dimensional Brans-Dicke theory.
The main difference with respect to the standard Brans-Dicke theory resides
in the fact that here the potential is now geometrically induced by the
fifth dimension in much the same way as the energy-momentum tensor comes
from the 5D pure vacuum. In the next section we shall give an application of
the above formalism in a cosmological context.

\section{The five-dimensional analogue of the O'Hanlon and Tupper solution}

A most known vacuum solution in Brans-Dicke theory of gravity is the
O'Hanlon and Tupper \cite{Tupper} solution. In this case $V(\phi )=0$ and
the range of values of parameter $\omega $ is restricted to $\omega >-3/2$, $%
\omega \neq 0,-4/3$ \cite{Faraoni}. In this section we shall obtain the
natural extension of this solution in five-dimensional Brans-Dicke theory..
We start by considering the metric corresponding to a homogeneous and
isotropic cosmological model in five-dimensional space%
\begin{equation}
dS^{2}=dt^{2}-a^{2}(t)[dx^{2}+dy^{2}+dz^{2}+dl^{2}],  \label{t1}
\end{equation}%
where $t$ is the cosmic time, $(x,y,z)$ are Cartesian coordinates and $l$ is
the fifth coordinate. Given that we are assuming homogeneity and isotropy we
should have $\varphi =\varphi (t)$, hence the field equations in vacuum (\ref%
{f6}) and (\ref{f7}) reduce to%
\begin{equation}
6H^{2}=\frac{\omega }{2}\frac{\dot{\varphi}^{2}}{\varphi ^{2}}+\frac{\ddot{%
\varphi}}{\varphi },  \label{t2}
\end{equation}%
\begin{equation}
3\frac{\ddot{a}}{a}+3H^{2}=-\frac{\omega }{2}\frac{\dot{\varphi}^{2}}{%
\varphi ^{2}}+H\frac{\dot{\varphi}}{\varphi },  \label{t3}
\end{equation}%
\begin{equation}
\ddot{\varphi}+4H\dot{\varphi}=0,  \label{t4}
\end{equation}%
where $H(t)=\dot{a}(t)/a(t)$ is the Hubble parameter. From equation (\ref{f8}%
) and taking into account that $^{(5)}R=-8\ddot{a}/a-12H^{2}$, we have 
\begin{equation}
\dot{H}=-\frac{\omega }{8}\left( \frac{\dot{\varphi}}{\varphi }\right) ^{2}-%
\frac{10}{4}H^{2}.  \label{t5}
\end{equation}%
From (\ref{t2}) and (\ref{t4}), the equation (\ref{t5}) becomes 
\begin{equation}
\dot{H}=-\frac{\omega }{3}\left( \frac{\dot{\varphi}}{\varphi }\right) ^{2}+%
\frac{5}{3}H\frac{\dot{\varphi}}{\varphi }.  \label{t6}
\end{equation}%
Let us try to obtain solutions with the following form%
\begin{equation}
a(t)=a_{0}\left( \frac{t}{t_{0}}\right) ^{q_{\pm }},\qquad \varphi
(t)=\varphi _{0}\left( \frac{t}{t_{0}}\right) ^{s_{\pm }},  \label{t7}
\end{equation}%
where $q$, $s$, $a_{0}$ and $\varphi _{0}$ are constants. Inserting (\ref{t7}%
) in (\ref{t2}), (\ref{t4}) and (\ref{t6}) gives 
\begin{equation}
q_{\pm }=\frac{2+2\omega \pm \sqrt{4+3\omega }}{2(5+4\omega )}=\frac{1}{%
5+4\omega }\left[ \omega +1\pm \sqrt{\frac{3\omega +4}{4}}\right] ,
\label{t8}
\end{equation}%
\begin{equation}
s_{\pm }=\frac{1\mp 2\sqrt{4+3\omega }}{4\omega +5},  \label{t9}
\end{equation}%
where $s$ and $q$ are algebraically related by the equation $s+4q=1$. On the
hypersurfaces $\Sigma _{0}$ $(l=l_{0})$ the equations (\ref{t2}) and (\ref%
{t3}) read 
\begin{eqnarray}
3H^{2} &=&\frac{8\pi }{\varphi }\rho ^{(BD)}+\frac{\omega }{2}\left( \frac{%
\dot{\varphi}}{\varphi }\right) ^{2}-3H\frac{\dot{\varphi}}{\varphi }
\label{t11} \\
2\frac{\ddot{a}}{a}+H^{2} &=&-\frac{8\pi }{\varphi }P^{(BD)}-\frac{\omega }{2%
}\left( \frac{\dot{\varphi}}{\varphi }\right) ^{2}-\frac{\ddot{\varphi}}{%
\varphi }-2H\frac{\dot{\varphi}}{\varphi },
\end{eqnarray}%
where 
\begin{equation}
\rho ^{(BD)}\equiv T^{(BD)}\,^{t}\,_{t}=T_{\varphi
}^{t}\,_{t}+T^{(IMT)}\,^{t}\,_{t}=-\frac{H}{8\pi }\left[ \dot{\varphi}%
+3H\varphi \right] ,  \label{t12}
\end{equation}%
\begin{equation}
P^{(BD)}=-T^{(BD)}\,^{i}\,_{i}=-\left( T_{\varphi
}\,^{i}\,_{i}+T^{(IMT)}\,^{i}\,_{i}\right) =\frac{1}{8\pi }\left[ H\dot{%
\varphi}+\varphi (\dot{H}+3H^{2})\right] .  \label{t13}
\end{equation}%
By employing (\ref{t7}) and the relation $s+4q=1$, the above equations
reduce to 
\begin{equation}
\rho ^{(BD)}=\frac{1}{8\pi }\frac{q}{t^{2}}(q-1)\varphi ,\qquad P^{(BD)}=-%
\frac{1}{8\pi }\frac{q^{2}}{t^{2}}\varphi .  \label{t14}
\end{equation}%
Clearly, in order to have a physical induced energy density $\rho ^{(BD)}>0$
and a negative induce pressure $P^{(BD)}<0$, the condition $q>1$ is
required. According to (\ref{t8}), this condition implies $-4/3<\omega <-5/4$%
, which satisfies the weak energy condition $\omega \geq -4/3$ \cite{PGO}.
The 4D induced equation of state is then 
\begin{equation}
P^{(BD)}=-\left( \frac{q}{q-1}\right) \rho ^{(BD)},  \label{t15}
\end{equation}%
which for $4.33<q<\infty $ gives $-1.3<P^{(BD)}/\rho ^{(BD)}<-1$. It turns
out that this is in agreement with the values of the interval that comes
from observational data: $-1.3<P/\rho <-0.7$ whose values describe the
present quintessential expansion of the universe \cite{Spergel}.

\section{Embedding the O'Hanlon and Tupper solution in five-dimensional
vacuum space}

Let us now drop the assumption of isotropy in all spatial dimensions and
consider a five-dimensional space with the following metric 
\begin{equation}
dS^{2}=dt^{2}-a^{2}(t)(dx^{2}+dy^{2}+dz^{2})-dl^{2},  \label{h1}
\end{equation}%
where, as previously, $t$ is the cosmic time, $(x,y,z)$ are Cartesian
coordinates, $l$ denotes the fifth coordinate,and $a(t)$ is a cosmological
scale factor. Assuming that $\varphi =\varphi (t,l)$ the field equations (%
\ref{f6}) and (\ref{f7}) now give 
\begin{equation}
3H^{2}=\frac{\omega }{2}\left( \frac{\dot{\varphi}}{\varphi }\right) ^{2}+%
\frac{\omega }{2}\left( \frac{\overset{\star }{\varphi }}{\varphi }\right)
^{2}+\frac{\ddot{\varphi}}{\varphi }  \label{h2}
\end{equation}%
\begin{equation}
2\frac{\ddot{a}}{a}+H^{2}=-\frac{\omega }{2}\left( \frac{\dot{\varphi}}{%
\varphi }\right) ^{2}+\frac{\omega }{2}\left( \frac{\overset{\star }{\varphi 
}}{\varphi }\right) ^{2}-H\frac{\dot{\varphi}}{\varphi }  \label{h3}
\end{equation}%
\begin{equation}
\frac{\ddot{a}}{a}+3H^{2}=-\frac{\omega }{2}\left( \frac{\dot{\varphi}}{%
\varphi }\right) ^{2}-\frac{\omega }{2}\left( \frac{\overset{\star }{\varphi 
}}{\varphi }\right) ^{2}-\frac{\overset{\star \star }{\varphi }}{\varphi }
\label{h4}
\end{equation}%
\begin{equation}
\ddot{\varphi}+3H\dot{\varphi}-\overset{\star \star }{\varphi }=0.
\label{h5}
\end{equation}%
Using (\ref{h5}), equation (\ref{h2}) can be rewritten as 
\begin{equation}
H^{2}=\frac{\omega }{6}\left( \frac{\dot{\varphi}}{\varphi }\right) ^{2}-H%
\frac{\dot{\varphi}}{\varphi }+\frac{\omega }{6}\left( \frac{\overset{\star }%
{\varphi }}{\varphi }\right) ^{2}+\frac{1}{3}\frac{\overset{\star \star }{%
\varphi }}{\varphi }.  \label{h6}
\end{equation}%
The trace equation (\ref{f8}), with $^{(5)}R=-(6\dot{H}+12H^{2})$ yields 
\begin{equation}
\dot{H}=-\frac{\omega }{6}\left( \frac{\dot{\varphi}}{\varphi }\right) ^{2}+%
\frac{\omega }{6}\left( \frac{\overset{\star }{\varphi }}{\varphi }\right)
^{2}-2H^{2}.  \label{h7}
\end{equation}%
Inserting (\ref{h5}) and (\ref{h6}) in (\ref{h7}), this expression becomes 
\begin{equation}
\dot{H}=-\frac{\omega }{2}\left( \frac{\dot{\varphi}}{\varphi }\right)
^{2}+2H\frac{\dot{\varphi}}{\varphi }-\frac{\omega }{6}\left( \frac{\overset{%
\star }{\varphi }}{\varphi }\right) ^{2}-\frac{2}{3}\frac{\overset{\star
\star }{\varphi }}{\varphi }.  \label{h8}
\end{equation}%
Now, the equations (\ref{h5}), (\ref{h6}) and (\ref{h8}) constitute an
independent system of differential equations. In order to find solutions to
this system we assume that the scalar field can be separated in the form $%
\varphi (t,l)=f(l)\phi (t)$. With this assumption the equations (\ref{h5}), (%
\ref{h6}) and (\ref{h8}) become 
\begin{equation}
H^{2}=\frac{\omega }{6}\left( \frac{\dot{\phi}}{\phi }\right) ^{2}-H\frac{%
\dot{\phi}}{\phi }+\frac{\omega }{6}\left( \frac{\overset{\star }{f}}{f}%
\right) ^{2}+\frac{1}{3}\frac{\overset{\star \star }{f}}{f},  \label{h9}
\end{equation}%
\begin{equation}
\dot{H}=-\frac{\omega }{2}\left( \frac{\dot{\phi}}{\phi }\right) ^{2}+2H%
\frac{\dot{\phi}}{\phi }-\frac{\omega }{6}\left( \frac{\overset{\star }{f}}{f%
}\right) ^{2}-\frac{2}{3}\frac{\overset{\star \star }{f}}{f},  \label{h10}
\end{equation}%
\begin{equation}
\frac{\ddot{\phi}}{\phi }+3H\frac{\dot{\phi}}{\phi }-\frac{\overset{\star
\star }{f}}{f}=0.  \label{h11}
\end{equation}%
In the particular case when $f(l)=1$ the system (\ref{h9}) to (\ref{h11})
reduces to 
\begin{eqnarray}
&&H^{2}=\frac{\omega }{6}\left( \frac{\dot{\phi}}{\phi }\right) ^{2}-H\frac{%
\dot{\phi}}{\phi },  \label{h14} \\
&&\dot{H}=-\frac{\omega }{2}\left( \frac{\dot{\phi}}{\phi }\right) ^{2}+2H%
\frac{\dot{\phi}}{\phi }, \\
&&\ddot{\phi}+3H\dot{\phi}=0.
\end{eqnarray}%
Again, if we assume that the solutions have the form (\ref{t7}) we readily
obtain 
\begin{eqnarray}
&&q_{\pm }=\frac{\omega }{3(\omega +1)\mp \sqrt{3(2\omega +3)}}=\frac{1}{%
3\omega +4}\left[ \omega +1\pm \sqrt{\frac{2\omega +3}{3}}\right] ,
\label{h16} \\
&&s_{\pm }=\frac{1\mp \sqrt{3(2\omega +3)}}{3\omega +4},
\end{eqnarray}%
where now $3q+s=1$. Note that this is the O'Hanlon and Tupper vacuum
solution in four dimensions. Therefore we conclude that the each leaf $%
\Sigma $ of the foliation defined by $l=const$\ corresponds to nothing more
than\ O'Hanlon and Tupper spacetime. Note, \textit{en passant}, that even
though they represent vacuum solutions, \ both the embedded and the ambient
spaces are not Ricci-flat.

To obtain the induced 4D potential we proceed as follows. Carrying out a
separation of variables\ in equation (\ref{h11}) yields 
\begin{equation}
\ddot{\phi}+3H(t)\dot{\phi}+|\alpha |\phi =0,  \label{h17}
\end{equation}
\begin{equation}  \label{h18}
\frac{d^{2}f}{dl^{2}}+|\alpha |f(l)=0,
\end{equation}
where $\alpha <0$ \ is a separation constant. Considering a constant Hubble
parameter $H=H_{0}$, the general solutions of (\ref{h17}) and (\ref{h18})
are, respectively 
\begin{equation}
\phi (t)=B_{1}e^{-\frac{3}{2}H_{0}\pm \frac{1}{2}\sqrt{9H_{0}^{2}-4|\alpha |}%
},  \label{h19'}
\end{equation}
\begin{equation}  \label{ult1}
f(l)=B_{2}e^{\pm i\sqrt{|\alpha |}},
\end{equation}
where $B_{1}$ and $B_{2}$ are integration constants. In the case of a
power-law expanding universe where the scale factor takes the form given in (%
\ref{t7}) the solution of (\ref{h18}) remains unaltered, while the general
solution of (\ref{h17}) now becomes 
\begin{equation}
\phi (t)=t^{-\nu }\left[ D_{1}\mathcal{J}_{\nu }(\sqrt{|\alpha |}\,t)+D_{2}%
\mathcal{Y}_{\nu }(\sqrt{|\alpha |}\,t)\right] ,
\end{equation}
$\mathcal{J}_{\nu }$ and $\mathcal{Y}_{\nu }$ denoting the first and second
kind Bessel functions, $\nu =(3q-1)/2$, and $D_{1}$, $D_{2}$ integration
constants. On the other hand, by using equation (\ref{potential}) for a
separable Brans--Dicke scalar field we obtain 
\begin{equation}
V[\phi ]=\omega \left[ \frac{\overset{\star }{f}^{2}}{f}\right] _{l_{0}}\phi
.  \label{h21}
\end{equation}%
Finally, inserting (\ref{ult1}) into (\ref{h21}) lead to 
\begin{equation}
V[\phi ]=(2\omega +3)|\alpha |\phi (t),  \label{h22}
\end{equation}%
where we have chosen the integration constant $B_{2}=-[(2\omega +3)/\omega
]e^{\mp i\sqrt{|\alpha |}\,l_{0}}$. This specification of $B_{2}$ allows us
to write equation (\ref{h17}) as 
\begin{equation}
\ddot{\phi}+3H(t)\dot{\phi}-\frac{1}{2\omega +3}\left[ \phi V^{\prime }(\phi
)-2V(\phi )\right] =0,  \label{h23}
\end{equation}%
with the prime $(^{\prime })$ denoting derivative with respect to the scalar
field $\phi $. On the hypersurfaces $\Sigma $ the equations (\ref{h2}) and (%
\ref{h3}) are%
\begin{eqnarray}
H^{2} &=&\frac{8\pi }{3\varphi }\rho ^{(BD)}+\frac{\omega }{6}\left( \frac{%
\dot{\varphi}}{\phi }\right) ^{2}-H\frac{\dot{\varphi}}{\varphi },
\label{h25} \\
2\frac{\ddot{a}}{a}+H^{2} &=&-\frac{8\pi }{\varphi }P^{(BD)}-\frac{\omega }{2%
}\left( \frac{\dot{\varphi}}{\varphi }\right) ^{2}-H\frac{\dot{\varphi}}{%
\varphi }-\frac{1}{\varphi }[\ddot{\varphi}+3H\dot{\varphi}],
\end{eqnarray}%
where $P^{(BD)}=-\rho ^{(BD)}=-\overset{\star \star }{\varphi }/(8\pi )$.

Therefore, we see that, in the present approach, it is possible to induce a
linear potential and an equation of state for vacuum in four dimensions.
This seems to be a remarkable result, although in a way it should be
expected since the metric (\ref{h1}) does not induce matter on the
hypersurfaces $\Sigma $. In other words, the induced energy-momentum tensor $%
T^{(IMT)}$ is null for this metric. However, in this case there is a
contribution from the scalar field that makes $T^{(BD)}$ nonzero. On the
other hand, it seems natural that we have obtained an equation of state that
describes vacuum, since we do not have matter, just the scalar field varying
with respect the fifth coordinate.

\section{Final remarks}

In this paper we have developed a procedure in which we regard our spacetime
as a hypersurface embedded in a five-dimensional space, solution of the 
Brans-Dicke vacuum\ field equations. The geometry as well as the
energy-momentum tensor that acts as source of the curvature of the
four-dimensional spacetime is determined by pure vacuum in five-dimensions,
an idea which goes back to the old Kaluza-Klein theory \cite{Kaluza-Klein}\
and that has been revived recently by the induced-matter theory\ or
non-compactified Kaluza-Klein gravity \cite{Wesson}. Since the Brans-Dicke
theory usually (but not always \cite{Barros}) reduces to general relativity
when $\omega \rightarrow \infty $, the formalism developed here is, in a
certain sense, a generalization of the induced-matter approach. There are,
however, from a geometrical point of view, significant differences. For
instance, in the induced-matter theory \textit{\`{a} la }Brans-Dicke\ the
Campbell-Magaard theorem cannot any longer be invoked to garantee the
embedding of any four-dimensional spacetime in five-dimensional Brans-Dicke
vacuum. This is because due to the presence of the Brans-Dicke scalar field
the ambient space is not, in general, Ricci-flat. However, it can be shown
that a new geometrical frame for five-dimensional Brans-Dicke theory exists
and is supported by an extension of the Campbell-Magaard theorem \cite%
{Dahia3}. On the other hand, as in the case of\ the induced-matter approach,
the theory provides no way of obtaining a unique spacetime from a given
five-dimensional metric, and that would require further mathematical
conditions on the embedding, or a kind of new dynamical principle to select
the possible choices of physically plausible spacetimes \cite{Stefan}.

\section*{Acknowledgements}

\noindent The authors acknowledge CNPq-CLAF and CNPq-FAPESQ (PRONEX) for
financial support.

\end{document}